\documentclass[twocolumn,superscriptaddress]{revtex4-2}
\usepackage[utf8]{inputenc}
\usepackage{times}
\usepackage[dvips]{graphicx}
\usepackage[usenames,dvipsnames]{xcolor}
\usepackage{amsmath}
\usepackage{amssymb}
\usepackage{bm}

\DeclareUnicodeCharacter{3000}{\textcolor{red}{BAD!!}}

\newcommand{\Revised}[1]{{\color{black}#1}}

\DeclareMathOperator{\erfc}{erfc}

\begin{document}

\title{Brownian yet non-Gaussian diffusion of a light particle in heavy gas: Lorentz gas based analysis}

\author{Fumiaki Nakai}
\email{nakai.fumiaki.c7@s.mail.nagoya-u.ac.jp}
\affiliation{Department of Materials Physics, Graduate School of Engineering, Nagoya University, Furo-cho, Chikusa, Nagoya 464-8603, Japan}

\author{Takashi Uneyama}
\affiliation{Department of Materials Physics, Graduate School of Engineering, Nagoya University, Furo-cho, Chikusa, Nagoya 464-8603, Japan}

\begin{abstract}
Non-Gaussian diffusion was recently observed in a gas mixture with mass and fraction contrast [F. Nakai et al, Phys. Rev. E {\bf 107}, 014605 (2023)].
The mean square displacement of a minor gas particle with a small mass is linear in time, while the displacement distribution deviates from the Gaussian distribution, which is called the Brownian yet non-Gaussian diffusion.
In this work, we theoretically analyze this case where the mass contrast is sufficiently large.
Major heavy particles can be interpreted as immobile obstacles, and a minor light particle behaves like a Lorentz gas particle within an intermediate time scale.
Despite the similarity between the gas mixture and the conventional Lorentz gas system, the Lorentz gas description cannot fully describe the Brownian yet non-Gaussian diffusion.
A successful description can be achieved through a \Revised{canonical} ensemble average of the statistical quantities of the Lorentz gas over the initial speed.
\end{abstract}
\maketitle

\section{Introduction}
\label{sec:introduction}

Gas diffusion is a classical problem \cite{chapman1990, jeans1921, mazenko2008, dorfman2021}, and it may be considered to be fully understood nowadays.
However, recent work revealed that gas diffusion is not that simple nor fully understood \cite{nakai2023fluctuating}.
The authors\cite{nakai2023fluctuating} numerically investigated the diffusion of a light gas particle in gas mixtures with mass and fraction contrast and found that the minor light molecule exhibits Brownin yet non-Gaussian diffusion: the mean square displacement (MSD) is linear in time \Revised{$\mathrm{MSD}\sim t^1$}, while the displacement distribution deviates from the Gaussian distribution.
\Revised{(We do not consider the anomalous diffusion $\mathrm{MSD}\sim t^{\delta}$ ($\delta\ne 1$ ) accompanied by a non-Gaussian displacement distribution, in what follows. Such diffusion is non-Brownian and non-Gaussian, while it is often observed in glassy liquids \cite{chaudhuri2007universal}, polymeric liquids\cite{ge2018nanorheology}, or some complex systems.)}
The Brownian yet non-Gaussian diffusion has been widely observed in complex systems with heterogeneous environments and/or conformational degrees of freedom, such as glass-forming liquids \Revised{\cite{rusciano2022fickian, chaudhuri2007universal, miotto2021length}}, polymeric fluids \cite{miyaguchi2017elucidating, uneyama2015fluctuation}, colloidal suspensions \cite{kim2013simulation, guan2014even}, \Revised{confined systems \cite{alexandre2023non}}, biological systems \cite{wang2009anomalous, he2016dynamic, jeon2016protein}, and active matters \cite{leptos2009dynamics, kurtuldu2011enhancement}.
In contrast to these complex systems, there is no heterogeneity nor internal degrees of freedom in the gas mixture.
In the previous work \cite{nakai2023fluctuating}, the origin of the non-Gaussian behavior was attributed to the fluctuating diffusivity which arises from a \Revised{separation of two relaxation timescales of the minor light particle velocity.
The relaxation timescale for speed (the magnitude of the velocity)
can be much longer than that for the direction of the velocity.
Namely, when the velocity is described in the spherical coordinates, the polar and azimuthal angle components rapidly relax at the time scale where the radial component almost remains unchanged.
}

The dynamics of a light particle in heavier particles have often been approximated as the Lorentz gas \cite{machta1983diffusion, andersen1964relaxation, dorfman2021, lebowitz1978transport}, which is composed of a mobile particle and immobile particles.
Although Lorentz gas was originally constructed for the dynamics of an electron in metal, later, the model has been regarded as the simple model for the transport phenomena of gas \cite{andersen1964relaxation} and also for some classical dynamical systems \cite{klages2019normal, zeitz2017active}.
Various transport properties of the Lorentz gas including the diffusion coefficient \cite{bruin1974computer, machta1983diffusion} or the relaxation of the velocity \cite{andersen1964relaxation, machta1983diffusion} are analyzed \Revised{even for dense cases \cite{hofling2006localization, hofling2007crossover, hofling2008critical, hofling2013anomalous}}.
Diverse extended models \Revised{\cite{klages2019normal, zeitz2017active, leitmann2017time, sanvee2022normal} including the experimental systems \cite{schnyder2017dynamic, siboni2018nonmonotonic, skinner2013localization}, glass forming liquids\cite{krakoviack2009tagged,voigtmann2009double}} have also been extensively studied.
Naively, we expect that the diffusion of a light gas particle in the heavy gas particles can be described using Lorentz gas.
To the best of the authors' knowledge, the relation between the simple dilute Lorentz gas and diffusion of a light particle in a matrix of heavy gas particles is not clear, and whether the non-Gaussian behavior can be described by the Lorentz gas is not clear, neither.

In the current work, we theoretically analyze the diffusion of a light gas particle in heavy gas particles.
We employ the dilute random Lorentz gas to describe the Brownian yet non-Gaussian diffusion of a light particle in heavy gas.
We first derive analytical expressions for the MSD and the non-Gaussian parameter of the Lorentz gas using the point process \cite{cox1980point}.
\Revised{(One can also calculate these correlation functions using the Boltzmann equation \cite{dufty2005brownian,brey1999brownian,garzo2019granular} and will reach the same results.)}
Afterward, we calculate the \Revised{canonical} ensemble average of them over the initial speed, which obeys the Maxwell-Boltzmann distribution.
The averaged result quantitatively reproduces the Brownian yet non-Gaussian diffusion of a light and minor gas particle in binary gas mixtures within an intermediate timescale.

\begin{figure}
    \centering
    \includegraphics[width=0.35\textwidth]{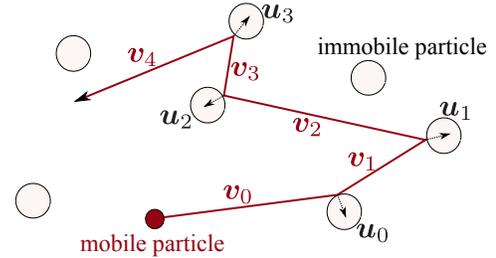}
    \caption{
Binary gas mixture as a Lorentz gas, which consists of a single mobile particle (point mass) in immobile spherical particles. $\bm{v}_i$ denotes the mobile particle velocity after $i$-th collision, and $\bm{u}_i$ is the unit vector from the mobile particle to the colliding immobile particle.}
    \label{fig:lorentz_gas}
\end{figure}

\section{Model}
\label{sec:model}

We consider a system that consists of the mobile particle with mass $m$ and size $0$ (point mass) in the fixed spherical obstacles with radius $\sigma$.
Note that we introduce the mass $m$ to analyze the binary gas mixture, although the conventional Lorentz gas model is independent of $m$.
We limit ourselves to the case where the fixed obstacles are dilute.
The fixed obstacles are randomly distributed in space, and there is no statistical correlation between different obstacles.
We describe the number density of the obstacles as $\rho$.
The mobile particle ballistically moves until it collides with an immobile obstacle.
When the mobile particle collides with the obstacle, the velocity of the mobile particle instantaneously changes via the hard-core repulsion potential \cite{dorfman2021}.
If we describe the velocities of the mobile particle before and after the $i$-th collision as $\bm{v}_{i}$ and $\bm{v}_{i + 1}$, they are related as
\begin{equation}
    \bm{v}_{i+1}=(\bm{1}-2\bm{u}_i \bm{u}_i) \cdot \bm{v}_{i}
    \label{eq:velocity_change}
\end{equation}
where $\bm{u}_i$ is the unit vector connecting the mobile particle to the colliding immobile particle, and $\bm{1}$ is the unit tensor.
See Fig.~\ref{fig:lorentz_gas} for a schematic representation of our model.
From Eq.~\eqref{eq:velocity_change}, the speed of the mobile particle $v$ remains unchanged: $v = |\bm{v}_i|$ (for any $i$).
\Revised{Here, we assume that the stochastic process of the mobile particle obeys the Markovian process; the successive collisions are uncorrelated.
This assumption is sufficient to describe the Brownian yet non-Gaussian diffusion that emerges in the binary gas mixture \cite{nakai2023fluctuating}.
(Due to this Markovian assumption, our model cannot reproduce so-called the
long-time tail, which is caused by correlated collisions \cite{alder1978long,brey1983long,ernst1971long,machta1984long}.)}
If we choose $m$, $v$, and $1/\rho\sigma^2$ to define the dimensionless units, there are no parameters in the random dilute Lorentz gas.

The model explained above can be interpreted as an approximate description for a minor and light particle in binary gas mixtures with sufficiently large mass and fraction contrast.
In such a binary gas mixture, the heavy particles are not entirely immobile but can move very slowly.
This means that the speed of the mobile particle is not exactly constant but just approximately constant at a specific time scale.
\Revised{In this sense, Eq.~\eqref{eq:velocity_change} is not exact
for a light particle but it should be interpreted as an approximation.}
Later, we will discuss a relation between the Lorentz gas and the \Revised{minor light molecule in the binary gas mixture.}

\section{Theory and Discussions}

\subsection{Lorentz Gas}
\label{subsec:lorentz_gas}

The dynamics of the mobile particle are described using $\bm{v}_i$ and $t_i$, where $t_i$ is the collision time at the $i$-th collision.
Using $\bm{v}_i$ and $t_i$, the position of the mobile particle at the time $t$ after the $n$-th collision, $\bm{r}(n,t)$, can be described as
\begin{equation}
 \bm{r}(n, t)=\Revised{(t-t_n)\bm{v}_n+\sum_{i=0}^{n-1} (t_{i+1}-t_i)\bm{v}_i}.
 \label{eq:position}
\end{equation}
To describe the dynamics, we require the collision statistics between successive collisions.
In the current case, the collision frequency density $f(\bm{u})$ for a collision with a direction vector $\bm{u}$ thoroughly characterizes the collision statistics.
$f(\bm{u})$ is obtained from the collision statistics for the binary gas mixture\cite{nakai2023fluctuating, mazenko2008} by setting the surrounding gas velocity to be $0$:
\begin{equation}
    f(\bm{u}_i)=\rho\sigma^2\bm{v}_i\cdot\bm{u}_i
    \Theta (\bm{v}_i\cdot\bm{u}_i),
    \label{eq:frequency_n}
\end{equation}
where $\Theta(x)$ is the Heaviside step function.
Eq.~\eqref{eq:frequency_n} can be rewritten into a simple form in the spherical coordinates.
Without loss of generality, we can take the Cartesian coordinates for $\bm{v}_{i}$ as $\bm{v}_i=(0,0,v)$ and express $\bm{u}_{i}$ as $\bm{u}_i=(\sin\theta_i\cos\phi_i, \sin\theta_i\sin\phi_i, \cos\theta_i)$ with $\theta_i\in [0, \pi/2)$ and $\phi_i\in[0, 2\pi]$.
Then Eq.~\eqref{eq:frequency_n} reduces to
\begin{equation}
    f(\bm{u}_i)=\rho\sigma^2 v\cos\theta_i
    \label{eq:frequency_n_reduce}
\end{equation}
By integrating Eq.~\eqref{eq:frequency_n_reduce} over $\bm{u}_{i}$, we obtain the collision frequency as
\begin{equation}
    \int f(\bm{u}_i) d\bm{u}_i = \frac{v}{\lambda}
    \label{eq:frequency}
\end{equation}
where $\lambda=1/\pi\rho\sigma^2$ is the mean free path.
Combining Eqs.~\eqref{eq:frequency_n_reduce} and  \eqref{eq:frequency}, we obtain the probability density where the mobile particle collides at $\bm{u}_i$ at the time interval $t_{i+1}-t_i$ for a given $v$, $P(t_{i+1}-t_i, \bm{u}_i; v)$ as \cite{nakai2023fluctuating}
\begin{equation}
    P(t_{i+1}-t_i, \bm{u}_i; v)=f(\bm{u}_i)\exp\left[-(t_{i+1}-t_i)\frac{v}{\lambda}\right].
    \label{eq:collision_statistics}
\end{equation}

The probability density at time $t$ with the number of collisions $n$ and the speed $v$ can be calculated using Eq.~\eqref{eq:collision_statistics}:
\begin{equation}
 \label{eq:collision_probability_full}
\begin{split}
    P(\bm{r}, \{\bm{u}_i \}, \{t_i\}; n, t, v)=&\delta[\bm{r}- \bm{r}(n, t)] \\
    &\times \prod_{i=0}^{n}P(t_{i+1}-t_i, \bm{u}_i, v).
\end{split}
\end{equation}
Integrating Eq.~\eqref{eq:collision_probability_full} over $\lbrace \bm{u}_{i} \rbrace$ and $\lbrace t_{i} \rbrace$, we have the probability density for $\bm{r}$ under given $n$, $t$, and $v$:
\begin{equation}
 \label{eq:collision_probability_reduced}
\begin{split}
    &P(\bm{r}; n, t, v)\\
    =&\int d\bm{v}_0
    \int_t^{\infty}dt_{n+1}
    \int_0^{t}dt_{n}
    \int_0^{t_n}dt_{n-1}
    \cdots
    \int_0^{t_2}dt_{1}\\
    &\times \int d\bm{u}_{n} \cdots \int d\bm{u}_{0}
    P(\bm{r}, \{\bm{u}_i \}, \{t_i\}; n, t, v)
    P(\bm{v}_0)\\
    =&e^{- vt/\lambda}
    \int d\bm{v}_0
    \int_0^{t}dt_{n}
    \int_0^{t_n}dt_{n-1}
    \cdots
    \int_0^{t_2}dt_{1}\\
    & \times \int d\bm{u}_{n-1} \cdots \int d\bm{u}_{0}
    \delta[\bm{r}-\bm{r}(n, t)]
    \prod_{i=0}^{n-1}f(\bm{u}_i)
    P(\bm{v}_0; v),
\end{split}
\end{equation}
where $P(\bm{v}_0; v) = \delta(|\bm{v}_{0}| - v) / 4 \pi v^{2}$ is the initial velocity distribution of the mobile particle.

To proceed with the calculation, we introduce the characteristic function, $C(\bm{k}; n, t, v)=\int d\bm{r} e^{i\bm{k}\cdot\bm{r}} P(\bm{r}; n, t, v)$.
From Eq.~\eqref{eq:collision_probability_reduced}, we can calculate $C(\bm{k}; n, t, v)$ as
\begin{equation}
\begin{split}
    &C(\bm{k}; n, t, v)\\
    =&e^{-vt/\lambda}
    \int d\bm{v}_0
    \int_0^{t}dt_{n}
    \int_0^{t_n}dt_{n-1}
    \cdots
    \int_0^{t_2}dt_{1}\\
    &\times \int d\bm{u}_{n-1} \cdots \int d\bm{u}_{0}
    \prod_{i=0}^{n-1} f(\bm{u}_i)
    P(\bm{v}_0;v)\\ &\times \exp\left\{i\bm{k}\cdot\left[\bm{v}_n(t-t_n)+\sum_{i=0}^{n-1}\bm{v}_i(t_{i+1}-t_i)\right]\right\} .
\end{split}
\label{eq:chacteristic_function}
\end{equation}
Here we consider the Laplace transform of Eq.~\eqref{eq:chacteristic_function}: $\hat{C}(\bm{k}; n, s, v) = \mathcal{L} [{C}(\bm{k}; n, \cdot, v)](s)
= \int_{0}^{\infty} dt \, e^{-t s} C(\bm{k}; n, t, v)$.
Then we have
\begin{equation}
\begin{split}
 \hat{C}(\bm{k}; n, s, v)
    = &\int d\bm{v}_{0}
    \frac{P(\bm{v}_0;v)}{s-i\bm{k}\cdot\bm{v}_{0}+v/\lambda} \\
 & \times \prod_{i=0}^{n-1} \int d\bm{u}_{i}
    \frac{f(\bm{u}_i)}{s-i\bm{k}\cdot\bm{v}_{i+1}+v/\lambda}.
\label{eq:chacteristic_function_laplace}
\end{split}
\end{equation}
The integrals over $\bm{v}_{0}$ and $\lbrace \bm{u}_{i} \rbrace$ in Eq~\eqref{eq:chacteristic_function_laplace}
can be analytically calculated. The result is
\begin{equation}
    \hat{C}(\bm{k}; n, s, v) 
    =\frac{\lambda}{v}\left[\frac{1}{k\lambda}
    \arctan\left(\frac{kv}{s+v/\lambda}\right)\right]^{n+1} ,
\label{eq:van_hove_result_n}
\end{equation}
where $k = |\bm{k}|$. To obtain the probability density for $\bm{r}$ under given $s$ and $v$, we need to consider all the contributions from different $n$.
This can be easily calculated by taking the summation over $n$: $\hat{C}(\bm{k};s,v) = \sum_{n = 0}^{\infty} \hat{C}(\bm{k};n,s,v)$.
From Eq.~\eqref{eq:van_hove_result_n}, we have
\begin{equation}
 \hat{C}(\bm{k};s,v)
    =\frac{\displaystyle 
    \arctan\left(\frac{kv}{s+v/\lambda}\right)}
 {\displaystyle (v / \lambda) \left[ k\lambda - 
    \arctan\left(\frac{kv}{s+v/\lambda}\right) \right]}.
    \label{eq:van_hove_ft_lt}
\end{equation}
Eq.~\eqref{eq:van_hove_ft_lt} corresponds to the Fourier-Laplace transform of the self part of the van Hove correlation function, and thus any quantities which characterize the diffusion behavior can be calculated from Eq.~\eqref{eq:van_hove_ft_lt}.
Note that Eq.~\eqref{eq:van_hove_ft_lt} satisfies the normalization condition of the probability density: $\hat{C}(\bm{k}; s,v) =s^{-1}$ at $k=0$.

The MSD is calculated as the second-order moment for the van Hove correlation function.
From Eq.~\eqref{eq:van_hove_ft_lt}, we have the Laplace transform of the MSD under a given $v$ as follows:
\begin{equation}
\begin{split}
    \mathcal{L}[\langle \bm{r}^2(\cdot)\rangle_{v}](s)
    =&-\left.\frac{\partial^2}{\partial\bm{k}^2} \hat{C}(\bm{k}; s, v)\right|_{\bm{k}=\bm{0}}\\
    =&\frac{2v^2}{s^2(s+v/\lambda)} ,
\end{split}
\label{eq:msd_lt}
\end{equation}
where $\langle\cdots\rangle_{v}$ denotes the statistical average under a given $v$.
Similarly, the Laplace transform of the fourth-order moment becomes
\begin{equation}
\begin{split}
    \mathcal{L}[\langle \bm{r}^4(\cdot)\rangle_{v}](s)
 =&\frac{\partial^2}{\partial\bm{k}^2}\frac{\partial^2}{\partial\bm{k}^2} \left. \hat{C}(\bm{k}; s, v) \right|_{\bm{k}=\bm{0}}\\
    =&\frac{8v^4(9s+5v/\lambda)}{3s^3(s+v/\lambda)^3} .
\end{split}
\label{eq:m4d_lt}
\end{equation}
The inverse Laplace transforms of Eqs.~\eqref{eq:msd_lt} and \eqref{eq:m4d_lt} give
\begin{align}
    \frac{\langle \bm{r}^2(t)\rangle_{v}}{\lambda^2}
    =&2 \left(-1+vt/\lambda+e^{-vt/\lambda}\right) ,
    \label{eq:msd_lorentz}\\
\begin{split}
    \frac{\langle \bm{r}^4(t)\rangle_{v}}{\lambda^4}
    =&
    \frac{4v^2t^2}{3\lambda^2}
    \left(5+4e^{-vt/\lambda}\right)-\\
    &\frac{8vt}{\lambda}
    \left(2-e^{-vt/\lambda}\right)
    +8\left(1-e^{-vt/\lambda}\right) .
\end{split}
    \label{eq:m4d_lorentz}
\end{align}
It is straightforward to show that at the short-time limit ($t \to 0$) Eqs.~\eqref{eq:msd_lorentz} and \eqref{eq:m4d_lorentz} approach $\langle \bm{r}^2(t)\rangle_{v} \to v^2t^2$ and $\langle \Revised{\bm{r}^4(t)}\rangle_{v} \to v^4t^4$.
These reflect the ballistic motion.
At the long-time limit $t\to \infty$, the MSD approaches $\langle \bm{r}^2(t)\rangle_{v} \to 2v\lambda t$, which corresponds to the normal diffusion.
The diffusion coefficient, $D$, defined as $\langle \bm{r}^2(t; v)\rangle=6Dt$, becomes $D=v\lambda/3$, which is consistent with the well-established result \cite{dorfman2021} in the gas kinetic theory.
From Eqs.~\eqref{eq:msd_lorentz} and \eqref{eq:m4d_lorentz}, the analytic expression of the non-Gaussian parameter (NGP) under a given $v$, $\alpha(t;v)$, is calculated to be
\begin{equation}
\begin{split}
    &\alpha(t; v)=\frac{3\langle \bm{r}^4(t)\rangle_{v}}
    {5\langle \bm{r}^2(t)\rangle_{v}^2}-1\\
    &=\frac{4e^{-vt/\lambda}(v^2t^2/\lambda^2-vt/\lambda+1)+1-2vt/\lambda-5e^{-2vt/\lambda}}
    {5(-1+vt/\lambda+e^{-vt/\lambda})^2}.
\end{split}
\label{eq:ngp_lorentz}
\end{equation}

\begin{figure}
    \centering
    \includegraphics[width=0.3\textwidth]{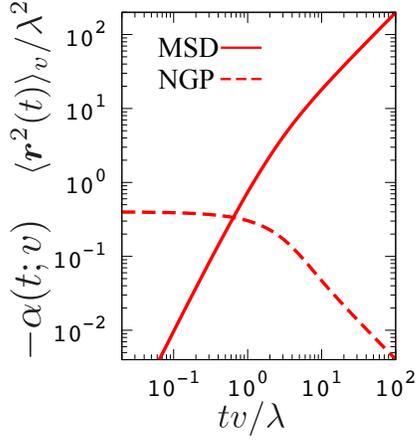}
    \caption{
    Theoretical predictions of the scaled mean square displacement (Eq.~\eqref{eq:msd_lorentz}) and non-Gaussian parameter (Eq.~\eqref{eq:ngp_lorentz}) against reduced time $v t / \lambda$ for the dilute random Lorentz gas.
}
    \label{fig:Lorentz-gas-msd-ngp}
\end{figure}

Fig.~\ref{fig:Lorentz-gas-msd-ngp} displays the MSD (Eq.~\eqref{eq:msd_lorentz}) and the absolute value of the NGP (Eq.~\eqref{eq:ngp_lorentz}) (the NGP by Eq.~\eqref{eq:ngp_lorentz} is always negative) of the Lorentz gas.
MSD shows simple ballistic and diffusive behaviors at the short and long timescales, respectively.
The crossover time is approximately equal to the mean free time, $vt/\lambda \approx 1$.
NGP becomes $-2/5$ at the short timescale and approaches $0$ at the long time scale.
The decay of $-\alpha(t;v)$ starts around $vt / \lambda \approx 1$, where the MSD switches from ballistic motion to normal diffusion.
The result that the NGP approaches zero at the long-time scale means that the dynamics of Lorentz gas can be reasonably described by the Gaussian process for $t \gtrsim \lambda / v$.
The non-Gaussianity at $t=0$ originates from the energy conservation of the Lorentz gas.
At the short timescale, the mobile particle ballistically moves, and thus NGP reflects the non-Gausianity of the velocity distribution.
We can easily evaluate the NGP at the short-time limit:
\begin{equation}
    \alpha(t;v) =\frac{3\langle \bm{v}^4t^4\rangle_{v}}{5\langle \bm{v}^2t^2\rangle_{v}^2}-1
    =-\frac{2}{5},
\end{equation}
which is consistent with the theoretical prediction at the short timescale in Fig.~\ref{fig:Lorentz-gas-msd-ngp}.

\Revised{
Before proceeding to the next section, we should mention the long-time tails, although they are beyond the scope of this work.
In general, molecules in matrices exhibit long-time correlated dynamics \cite{alder1978long,brey1983long,ernst1971long,machta1984long} and exhibit long-time tails at a long-time scale.
This means that the history of collisions (or called ring collisions) can not be strictly ignored even in any long timescale.
Such a correlation leads to some characteristic power laws in time correlation functions\cite{ernst1971long, machta1984long}.
The current system does not show such long-time tails because of the Markovian process assumption.
}

\subsection{Binary Gas Mixture}
\label{subsec:binary_gas_mixture}

The Lorentz gas has often been regarded as the model that describes a light particle in heavy particles.
Thus, one may expect that the Brownian yet non-Gaussian diffusion, which is observed for a minor light particle in heavy gas \cite{nakai2023fluctuating}, can be predicted from the Lorentz gas model.
However, the theoretical result of the random dilute Lorentz gas (Fig.~\ref{fig:Lorentz-gas-msd-ngp}) does not exhibit the non-Gaussian diffusion in the normal diffusion regime (MSD$\propto t$).

\begin{figure}
    \centering
    \includegraphics[width=0.3\textwidth]{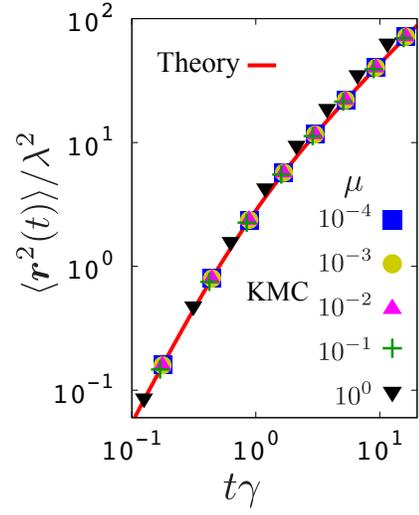}
    \caption{
Theoretical prediction for MSD of binary gas mixtures with different mass ratios.
The red curve is the theoretical prediction by Eq.~\eqref{eq:msd_ensemble}, and symbols are the results of KMC simulations\cite{nakai2023fluctuating}.
}
    \label{fig:msd_ensemble}
\end{figure}

\Revised{
This apparent inconsistency between the Lorentz gas model and a binary gas mixture comes from the fact that the speed (or kinetic energy) of a light gas particle in a binary gas mixture fluctuates at a very long timescale.
(We may interpret the results based on the Lorentz gas model are for the microcanonical ensemble, and we should use the canonical ensemble for the binary gas mixture.)
Within a certain timescale shorter than the relaxation timescale of the speed of the light particle, we can interpret the self-part of the van-Hove correlation function of the light particle in the gas mixture $G(\bm{r}, t)$ as the canonical ensemble average of that for the Lorentz gas model.
Using Eq.~\eqref{eq:collision_probability_reduced}, $G(\bm{r}, t)$ is described as
\begin{equation}
    G(\bm{r}, t)= \sum_{n=0}^\infty \int dv P(\bm{r}; n, t, v) P_{\text{MB}}(v),
    \label{eq:van-Hove_full}
\end{equation}
where $P_{\text{MB}}(v)$ is the Maxwell-Boltzmann distribution for the speed:
\begin{equation}
    P(v)=4\pi v^2\left(\frac{m}{2\pi k_BT}\right)^{3/2}
    \exp\left(-\frac{mv^2}{2k_BT}\right).
    \label{eq:Maxwell-Boltzmann}
\end{equation}
Here, $k_B$ is the Boltzmann constant.
While the calculation of Eq.~\eqref{eq:van-Hove_full} itself is difficult,
some moments for the displacement can be analytically obtained.
From Eqs.~\eqref{eq:msd_lorentz} and \eqref{eq:Maxwell-Boltzmann}, the second moment of $G(\bm{r}, t)$ is
}
\begin{equation}
\begin{split}
    \frac{\langle \bm{r}^2(t)\rangle}{\lambda^2}
    =& \int \frac{\langle \bm{r}^2(t) \rangle_{v}}{\lambda^2} P(v) dv\\
    =&2
    \left[-1+\frac{2\gamma t}{\sqrt{\pi}}
    +(1+2\gamma^2t^2)e^{\gamma^2t^2}\erfc(\gamma t)\right],
\end{split}
\label{eq:msd_ensemble}
\end{equation}
where $\gamma$ is a characteristic frequency defined as $\gamma=\sqrt{k_BT/2m}/\lambda$.
Fig.~\ref{fig:msd_ensemble} displays the prediction for the MSD of a light particle in a binary gas mixture by Eq.\eqref{eq:msd_ensemble}.
For comparison, the mean square displacements of the light particle in heavier particles, calculated from the kinetic Monte Carlo (KMC) simulations \cite{nakai2023fluctuating}, are also presented with various mass ratios \Revised{$\mu=m/M$ where $M$ is the mass of the major gas particle}.
Our theoretical prediction quantitatively agrees with the MSD from the KMC simulations when $\mu$ is sufficiently small ($\mu\ll 1$).

\begin{figure}
    \centering
    \includegraphics[width=0.4\textwidth]{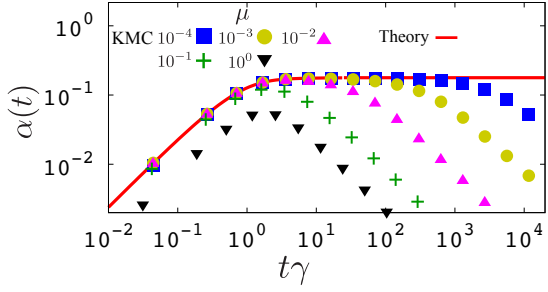}
    \caption{
Theoretical prediction for NGP of binary gas mixtures with different mass ratios.
The red curve is the theoretical prediction by Eqs.~\eqref{eq:msd_ensemble} and \eqref{eq:m4d_ensemble}, and symbols are the results from the KMC simulations\cite{nakai2023fluctuating}.
}
    \label{fig:ngp_ensemble}
\end{figure}

In a similar manner, the NGP for a binary gas mixture can be analytically calculated.
The ensemble average of the fourth-order moment is obtained using Eqs.~\eqref{eq:m4d_lorentz} and \eqref{eq:Maxwell-Boltzmann} as
\begin{equation}
\begin{split}
    &\frac{\langle \bm{r}^4(t)\rangle}{\lambda^4}
    =\int \frac{\langle \bm{r}^4(t)\rangle_{v}}{\lambda^4} P(v) dv\\
    =&\frac{4}{3}
    \left[
    6-\frac{12\gamma t}{\sqrt{\pi}}+30\gamma^2t^2
    -\frac{56\gamma^3t^3}{\sqrt{\pi}}
    -\frac{32\gamma^5t^5}{\sqrt{\pi}} \right.\\
    &\left. - (6+24\gamma^2t^2-72\gamma^4t^4-32\gamma^6t^6)
    e^{\gamma^2t^2}
    \erfc(\gamma t)
    \right] .
\end{split}
\label{eq:m4d_ensemble}
\end{equation}
Fig.~\ref{fig:ngp_ensemble} displays the NGP $\alpha(t)$ calculated from Eqs.~\eqref{eq:msd_ensemble} and \eqref{eq:m4d_ensemble}.
For comparison, the NGPs from the KMC simulations \cite{nakai2023fluctuating} are also shown with various $\mu$.
Our theoretical prediction successfully describes the non-Gaussian parameter from the KMC simulation for sufficiently small $\mu$ except for the decay of the NGP at the very long time scale.
We will discuss this discrepancy later.
The NGP by Eqs.~\eqref{eq:msd_ensemble} and \eqref{eq:m4d_ensemble} approaches $0$ at $t \to 0$, unlike the case of the Lorentz gas.
This reflects the fact that the probability density of the speed of the mobile particle obeys the Maxwell-Boltzmann distribution, which is a Gaussian distribution.
At the long-time limit ($t\to\infty$), the NGP approaches $3\pi/8-1$, and this quantity corresponds to the plateau of the NGP from the KMC simulations.

At last, we discuss the self part of the van Hove correlation function $G_{s}(\bm{r}, t)$ at the limit of the long time scale which is sufficiently larger than the mean free time, based on the Lorentz gas model.
At the long-time limit in the Lorentz gas ($s\ll v/\lambda$), Eq.~\eqref{eq:van_hove_ft_lt} reduces to
\begin{equation}
     \hat{C}(\bm{k};s,v)
    \approx \frac{\displaystyle 
    \arctan(k\lambda)-\frac{ks}{k^2v+v/\lambda^2}}
 {\displaystyle (v / \lambda) \left[ k\lambda - 
    \arctan(k\lambda)+\frac{ks}{k^2v+v/\lambda^2}\right]}.
\end{equation}
Its inverse-Laplace-transform is
\begin{equation}
\begin{split}
    &C(\bm{k};t,v)\\
    \approx    &(1+k^2\lambda^2)\exp\left[-\frac{(1+k^2\lambda^2)[k\lambda-\arctan(k\lambda)]}{k\lambda}\frac{tv}{\lambda}\right].
    \label{eq:van-hove_ft_limit}
\end{split}
\end{equation}
Here, we should note that Eq.~\eqref{eq:van-hove_ft_limit} is valid for the long-time scale, $t\gg \lambda/v$. In this timescale, we expect that only the small wavelength component ($k\lambda\ll 1$) becomes dominant.
Then Eq.\eqref{eq:van-hove_ft_limit} can be reduced to 
\begin{equation}
    C(\bm{k};t,v)     \approx  \exp\left(-\frac{v t\lambda k^2}{3}\right).
     \label{eq:van-hove_gaussian_limit}
\end{equation}
Eq.~\eqref{eq:van-hove_gaussian_limit} is nothing but the characteristic function of the Gaussian distribution.
Thus, the van-Hove correlation function of the Lorentz gas for a given speed $v$ at the long-time limit is
\begin{equation}
    G(x; t, v)     \approx  \sqrt{\frac{3}{4\pi v\lambda t}}\exp\left(-\frac{3 x^2}{4v \lambda t}\right) .
    \label{eq:van_hove_v}
\end{equation}
From Eqs.~\eqref{eq:Maxwell-Boltzmann} and \eqref{eq:van_hove_v}, we obtain the van-Hove correlation function for a light gas particle in a binary gas mixture as
\begin{equation}
\begin{split}
    G(x;t) = &\int_0^{\infty} G(x; t, v)P(v)dv,
\end{split}
    \label{eq:van_hove_approximation_ensemble}
\end{equation}
\Revised{
Although Eq.~\eqref{eq:van_hove_approximation_ensemble} can be analytically calculated, the result is rather complex.
Because the Brownian yet non-Gaussian diffusion often has striking features in tails in the van-Hove correlation function \cite{sposini2018random, jeon2016protein, Chechkin2017}, we attempt to obtain the approximate form for the tail for the large displacement region.
The saddle point approximation for Eq.~\eqref{eq:van_hove_approximation_ensemble} gives
\begin{equation}
\begin{split}
    G(x; t, v)     \approx
    &\sqrt{\frac{3m}{4\pi k_BT}}\frac{|x|}{\lambda t}
    \exp\left[-3\left(\sqrt{\frac{9m}{128k_BT}}\frac{x^2}{\lambda t} \right)^{\frac{2}{3}}\right]\\
    &\mathrm{for} \ x^2\gg \lambda t\sqrt{k_BT/m}) .
\end{split}
\label{eq:van-hove_tail}
\end{equation}
Eq.~\eqref{eq:van-hove_tail} is the same as the form obtained phenomenologically in our previous work \cite{nakai2023fluctuating}.
It would be worth stressing here that Eq.~\eqref{eq:van-hove_tail} is not exponential nor stretched Gaussian distributions.
}

As we mentioned, our theoretical prediction for the NGP $\alpha(t)$
converges to a constant value at the long-time limit.
Therefore, at least in the theoretical framework shown above, our theory does not predict the Gaussian behavior at the long-time region observed in the KMC simulations.
The discrepancy between the theoretical prediction and the KMC data at the long-time limit can be attributed to the lack of speed relaxation in our analysis.
The KMC simulations revealed that there are two characteristic relaxation time scales in a binary gas mixture: the direction and speed relaxations \cite{nakai2023fluctuating}.
In our analysis, we considered the direction relaxation via hard-core collisions, while the speed relaxation is not explicitly considered.
We only assumed that the speed relaxation makes the initial ensemble with the Maxwell-Boltzmann distribution.
The speed relaxation affects the long-time dynamics, but it is totally ignored.
Therefore, the diffusion coefficient for the particle with a given initial speed $v$ remains constant even at the long-time limit: $D = v \lambda / 3$.
This means that the long-time diffusion behavior reflects the initial speed distribution. This is why the NGP does not approach zero even at the long-time limit.

The diffusion coefficient of a light particle in a binary gas mixture fluctuates in time due to speed relaxation.
Therefore, if we describe the diffusion of the light particle at the long timescale, we need to incorporate another stochastic process into the model.
The diffusing diffusivity model \cite{Chechkin2017} in which the diffusion coefficient obeys the Langevin equation would be employed to incorporate the fluctuation of the speed.
Some analytical results of the gas kinetic theory \cite{pagitsas1979kinetic} can be utilized to design the stochastic process for the speed of the particle.

\section{conclusion}
In this work, we theoretically analyzed the dynamics of a light particle in heavy gas particles with large mass contrast, based on the dilute Lorentz gas model.
\Revised{
We derived the analytical expressions for the MSD and NGP of the Lorentz gas for a given speed $v$ via the point process approach.
The result with constant $v$ does not exhibit the Brownian yet non-Gaussian diffusion observed in the binary gas mixtures with mass and fraction contrast.
We found that the Brownian yet non-Gaussian diffusion can be reproduced through the canonical ensemble average of statistical quantities for the Lorentz gas over the initial speed}, except for the very long-time region.
This work revealed a relation between the conventional Lorentz gas and the binary gas mixture, and it will provide fresh insight into the theoretical modeling for gas diffusion.

\section*{Acknowledgments}
FN was supported by Grant-in-Aid for JSPS (Japan Society for the Promotion of Science) Fellows (Grant No. JP21J21725).

\appendix
\setcounter{figure}{5}

\Revised{
\section{Kinetic Monte Carlo simulation}\label{appendix_kmc}
We briefly explain the kinetic Monte Carlo simulation for the dynamics of a minor particle in the binary gas mixture with fraction contrast (the details are shown in Ref.~\cite{nakai2023fluctuating}).
In this method, we employ the following assumptions:
\begin{enumerate}
    \item \label{hardcore_assumption}
	  The minor particles interact with the major particle via the hard-core potential.
    \item \label{dilute_assumption}
	  The minor particles are infinitely diluted; the minor particles interact only with the major ones.
    \item \label{markovian_assumption}
	  The dynamics of the minor particle obey the Markovian process.
    \item \label{equilibrium_assumption}
	  The system is in an equilibrium state; the positions of the particles are homogeneously distributed, and the statistics of the velocity obey the Maxwell-Boltzmann distributions.
\end{enumerate}
These assumptions have often been employed in gas kinetics \cite{dorfman2021,mclennan1989introduction}.
From Assumption~\ref{hardcore_assumption}, the minor particle velocity $\bm{v}$ changes to $\bm{v}'$ by a collision with a major one as
\begin{equation}
    \bm{v}'=\bm{v}-\frac{2M}{m+M}(\bm{v}-\bm{V})\cdot\hat{\bm{r}}\hat{\bm{r}},
    \label{eq:appendix_velocity_change}
\end{equation}
where $\hat{\bm{r}}$ is the direction unit vector connecting the center of the minor particle to that of the colliding major particle and $\bm{V}$ is the velocity of the colliding major particle.
On the basis of Assumptions \ref{dilute_assumption}-\ref{equilibrium_assumption}, we can calculate
the collision statistics. The probability density where a minor particle collides with a major particle having $\bm{V}$ at $\hat{\bm{r}}$ and at the time interval $\tau$, for a given minor particle's velocity $\bm{v}$ is
\begin{equation}
\begin{split}
    &P(\bm{V}, \hat{\bm{r}}, s | \bm{v})\\
    & = \rho\sigma^2(\bm{v}-\bm{V})\cdot \hat{\bm{r}}
    \left(\frac{M}{2\pi k_BT}\right)^{3/2}
    \exp\left( -\frac{MV^2}{2k_BT} \right)\\
    &\times\exp\left[-F(v) s\right]
    \Theta[(\bm{v}-\bm{V})\cdot\hat{\bm{r}}],
\end{split}
\label{eq:appendix_collision_statistics}
\end{equation}
where $\Theta(x)$ is the Heaviside function and $F(v)$ is the collision frequency of the minor particle for a given $v$ \cite{nakai2023fluctuating}:
\begin{equation}
    F(v)=\pi\rho\sigma^2\left[
    \left(v+\frac{\xi^2}{2v}\right)\mathrm{erf}\left(\frac{v}{\xi}\right)
    +\frac{\xi}{\sqrt{\pi}}\exp\left(-\frac{v^2}{\xi^2}\right)
    \right],
\label{eq_appendix_collision_frequency}
\end{equation}
where we defined the characteristic major particle speed $\xi\equiv \sqrt{2k_BT/M}$.
The stochastic process of the minor particle is 
fully characterized by Eqs.~\eqref{eq:appendix_velocity_change}
and \eqref{eq:appendix_collision_statistics}.
At the large the mass ratio case ($\mu=m/M\ll 1$), which is the interest of this work, Eqs.~\eqref{eq:appendix_velocity_change} and \eqref{eq:appendix_collision_statistics} reduce to Eqs.~\eqref{eq:velocity_change} and \eqref{eq:frequency_n}.
}

\bibliographystyle{apsrev4-2}
\bibliography{ref}

\end{document}